\documentclass[onecolumn]{article}

\usepackage[paperwidth=8.5in, paperheight=11in]
    {geometry}
\usepackage{fullpage}
\usepackage{changepage}
\usepackage{amsmath}
\usepackage{graphicx}
\usepackage{morefloats}
\usepackage{verbatim}
\usepackage{pdflscape}
\usepackage{ifthen}
\usepackage{verbatim}
\usepackage{caption}
\usepackage{subcaption}
\usepackage[colorlinks=false]{hyperref}
\usepackage[all]{hypcap}

\title{Asymptotic behaviours of Unitary $9j$ Coefficients - Numerical Study}
\author{$^1$Brian Kleszyk, $^1$Larry Zamick, $^2$Ben Bayman \\\\
$^1$Department of Physics and Astronomy, Rutgers University, Piscataway, New Jersey 08854 \\ 
$^2$School of Physics and Astronomy, University of Minnesota, Minneapolis, Minnesota 55455
}

\date{\today}

\begin{document}
% % % % % % % % % % % % % % % % % % % % % ABSTRACT
\maketitle
\begin{abstract}
	We consider the asymptotic behaviors of selected unitary $9j$ coefficients. 
	We had examples of exponential decreases and of power law behaviors.
	We find that extreme precision is required in calculating these coefficients and that Mathematica can be used provided fractional rather than decimal input is used.
\end{abstract}

% % % % % % % % % % % % % % % % % % % % % INTRODUCTION
\section{Introduction}
	
	We here consider the asymptotic behaviour of certain unitary $9j$ coefficients. 
	These are defined in a previous work. 
	In particular we consider the following $U9j$'s:
	\begin{equation}
	U(I,j) = {\Big<(jj)^{2j} (jj)^{2j} \Big| (jj)^{2j} (jj)^{(2j - 2)} \Big>}^I 
	\end{equation}
	 with all even $I$ from 2 to 32.
	Previously the case $I=2$ was considered in \cite{1} but the largest $j$ considered was $j=100.5$.
	Here to insure we are truly in the asymptotic region we extend this to $j=7000.5$.
	Also we improve the method of analysis.
	Part of the motivation for this work comes from \cite{2} where we found that some coupling matrix elements were surprisingly small.
	We also consider cases where $I=I_{\text{max}} =4j-2$, $I_{\text{max}} -2$, e.t.c.
	
	We are dealing with numbers which range from one to exceedingly small values e.g. $10^{-500}$ so it is necessary to have very high precision in the calculations of $9j$ coefficients.
	We have found that Wolfram's Mathematica fits the bill provided it is used correctly.
	One must enter numbers as fractions rather than decimals e.g. 1001/2 rather than 500.5.

% % % % % % % % % % % % % % % % % % % % % CALCULATION
\section{Calculation}
	As was noted in \cite{1} at first glance $U(2,j)$ seems to fall of exponentially with $j$.
	This suggests a form 
	\begin{equation}
		A e^{-\alpha j}
	\end{equation}
	For this form $\ln(|U(2,j)|) = \ln(A)-\alpha j$.
	If this were true there would be a linear relationship between $\ln(|U(2,j)|)$ and $j$.
	We will here also consider other values of $I$ as indicated above.

\begin{figure}
\centering
   \begin{subfigure}{0.35\textheight} \centering
     \includegraphics[width=\textwidth]{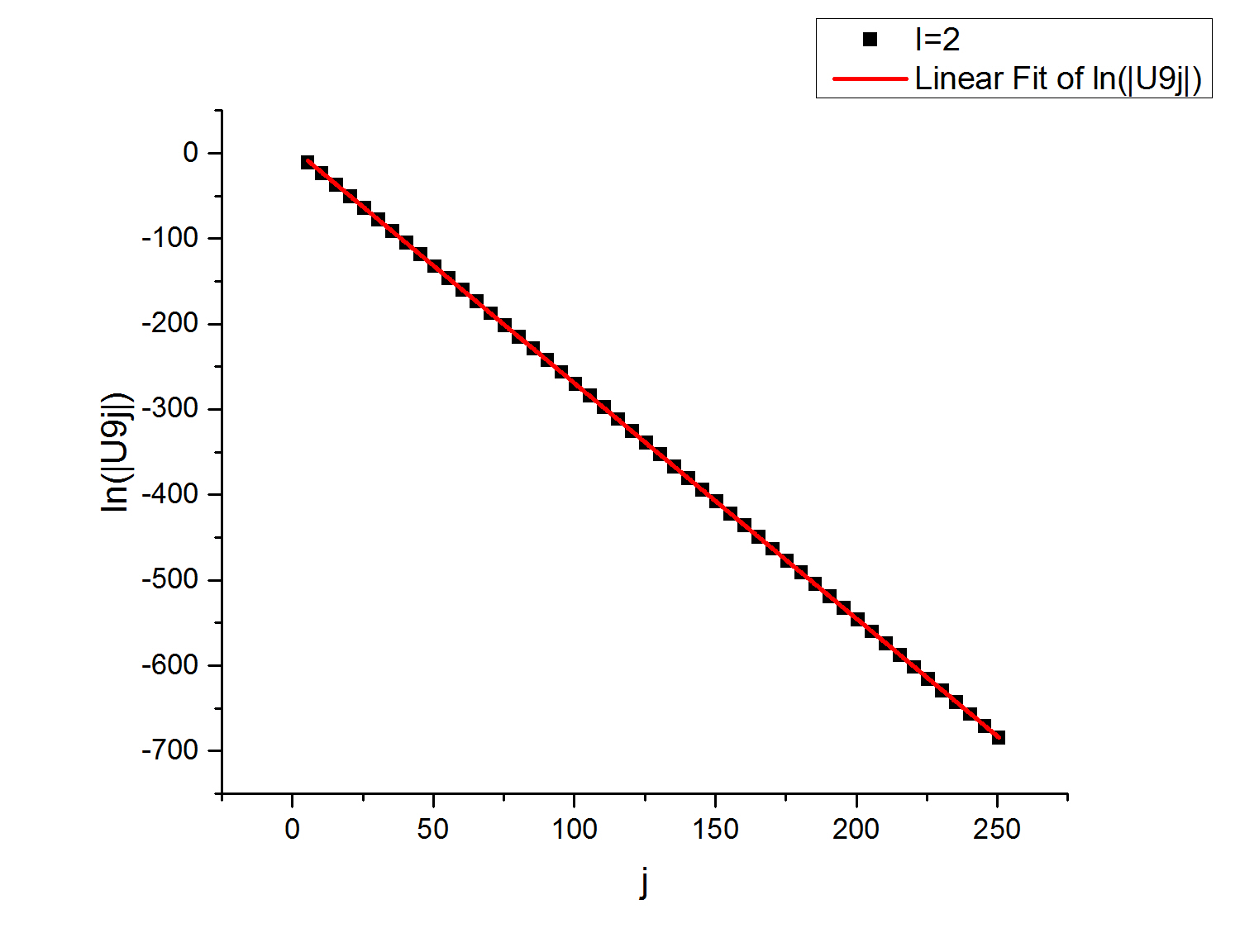}
     \caption{$\ln(|U9j|)$ vs $j$}
   \end{subfigure}
   \begin{subfigure}{0.35\textheight} \centering
     \includegraphics[width=\textwidth]{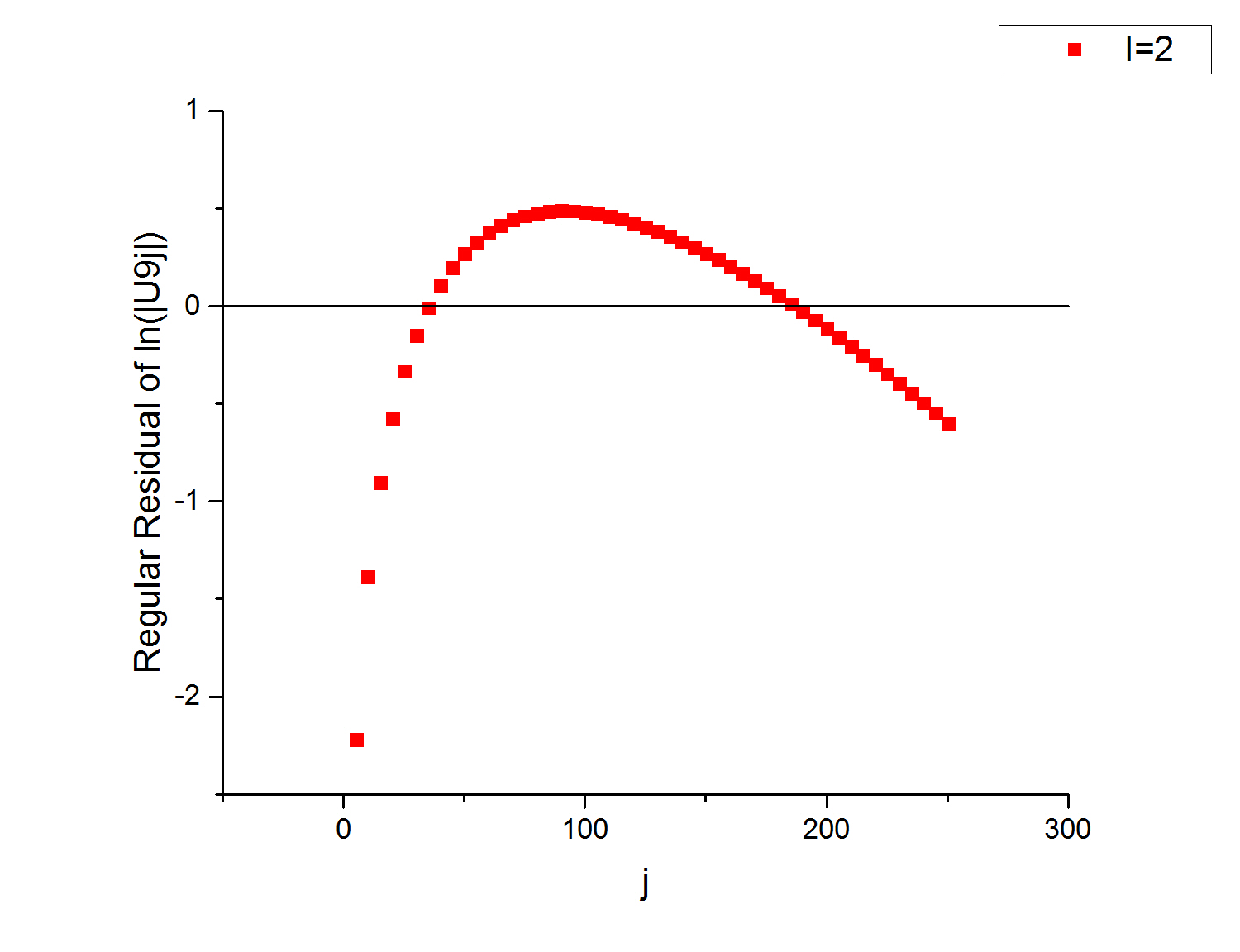}
     \caption{Residual of $\ln(|U9j|)$ vs $j$}
   \end{subfigure}
\caption{$I=2$ \label{fig:firstExp}} 
\end{figure}

\begin{figure}
\centering
   \begin{subfigure}{0.35\textheight} \centering
     \includegraphics[width=\textwidth]{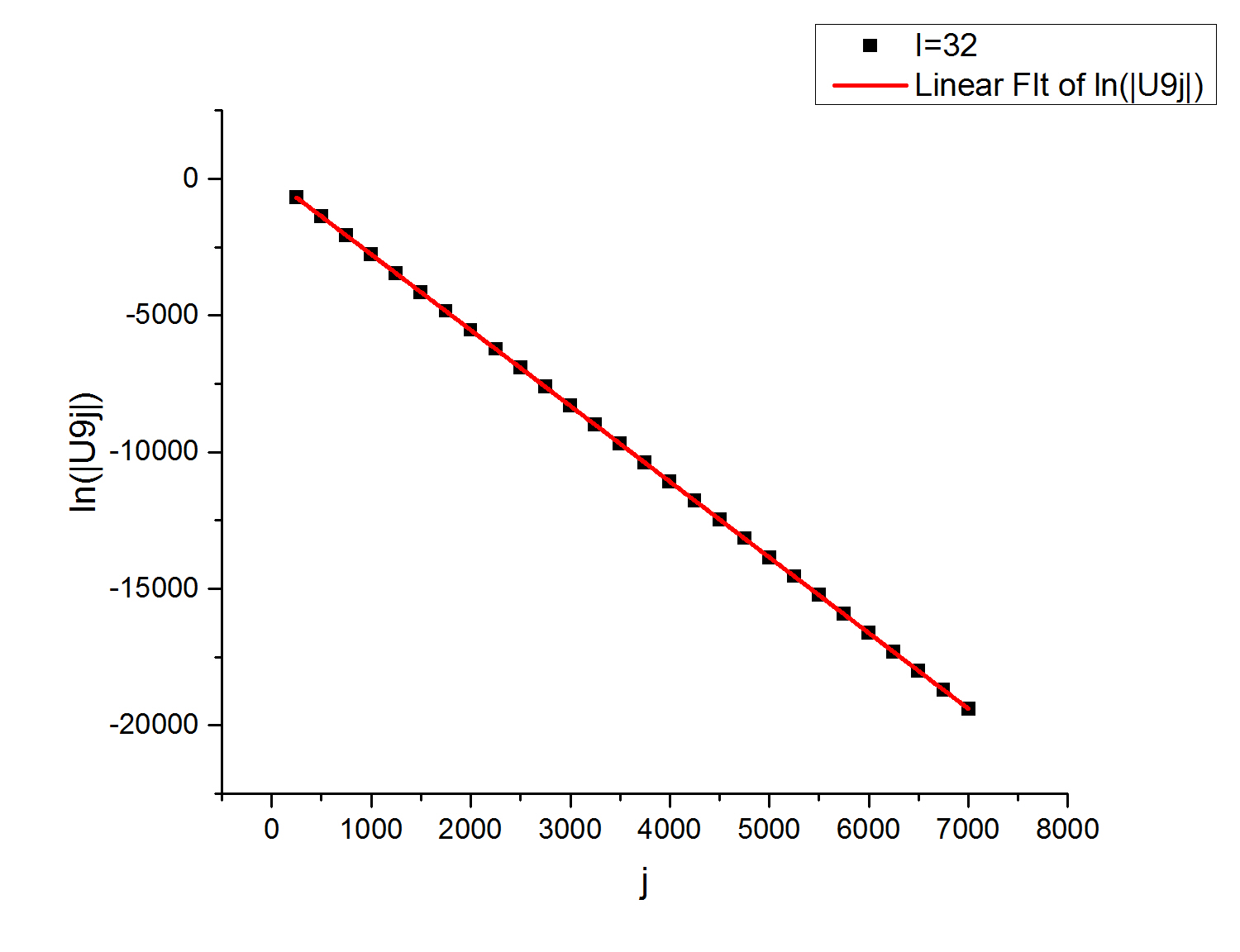}
     \caption{$\ln(|U9j|)$ vs $j$}
   \end{subfigure}
   \begin{subfigure}{0.35\textheight} \centering
     \includegraphics[width=\textwidth]{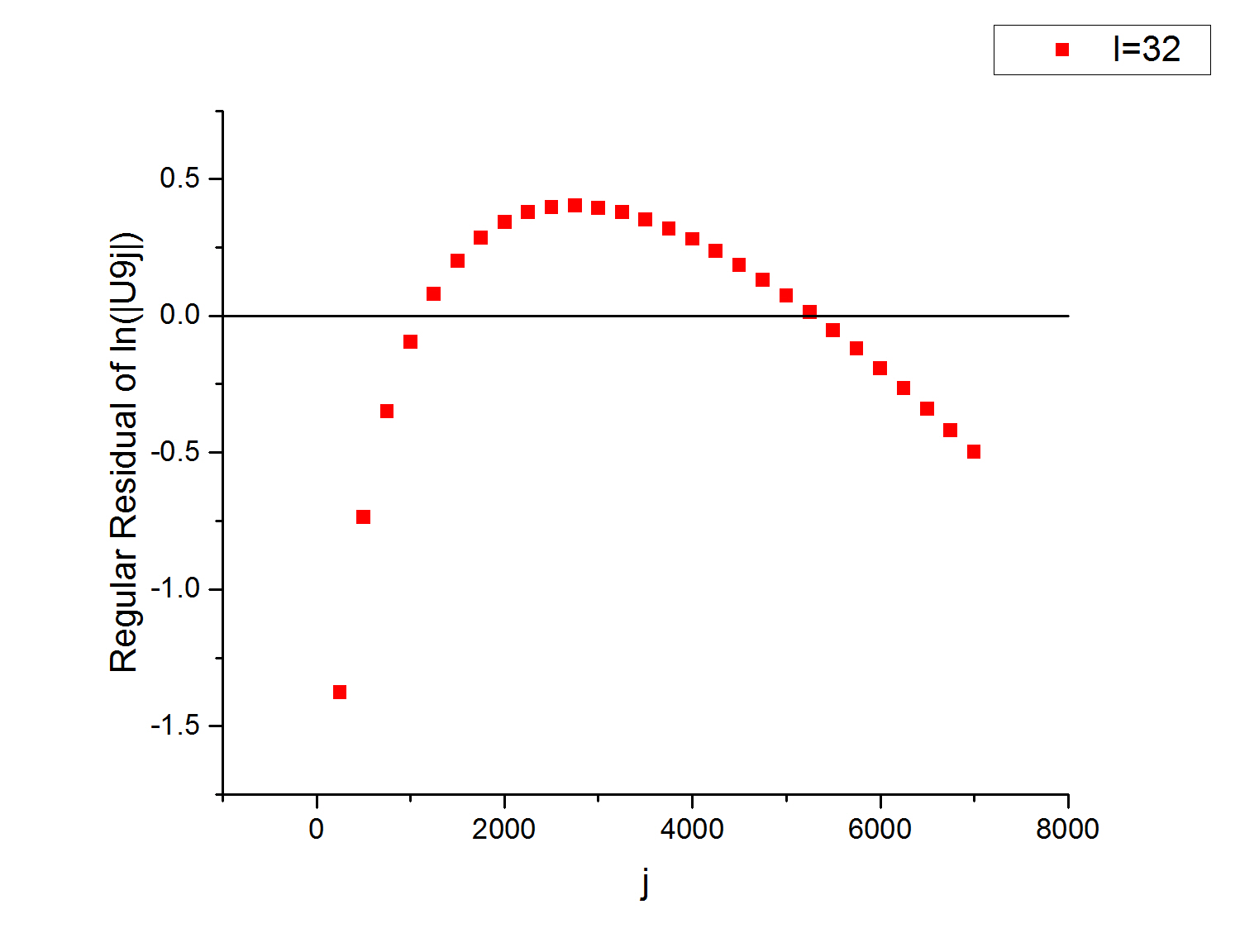}
     \caption{Residual of $\ln(|U9j|)$ vs $j$}
   \end{subfigure}
\caption{$I=32$\label{fig:lastExp}} 
\end{figure}

	We first plot, in Figures~\ref{fig:firstExp} to~\ref{fig:lastExp}, $\ln(|U(I,j)|)$ vs $j$ for some even $I$ values between $I=2$ and $I=32$. 
	The curves indeed looks like straight lines indicating that the $U(I,j)$'s drops exponentially with $j$.
	This is certainly the dominant trend but there are small deviations indicated by the error analysis. 

\begin{figure}
     \centering
	\parbox{0.48\textwidth}{ \centering
		\includegraphics[width=.48\textwidth]
		  {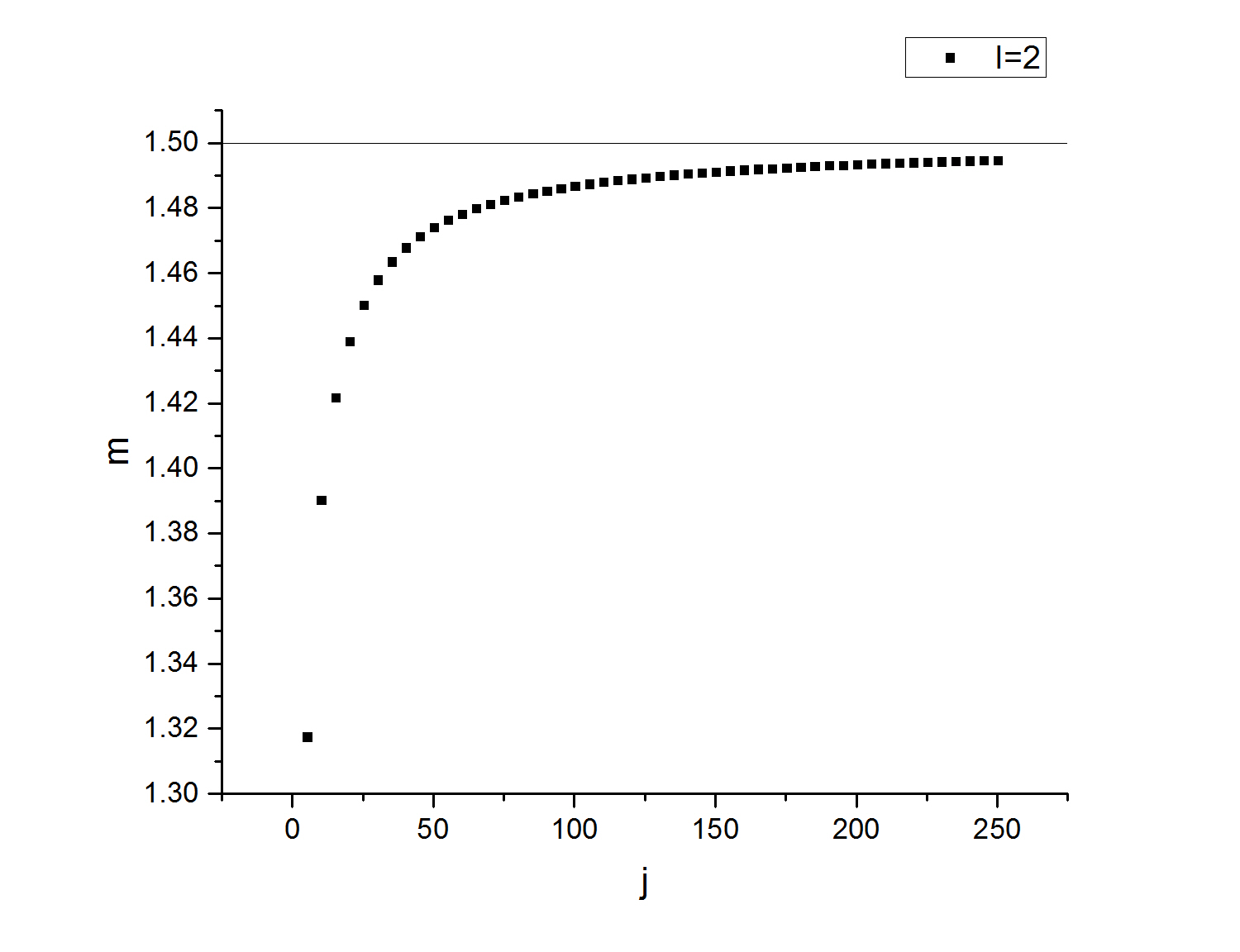}
		\caption{Suspected $m$ vs $j$; $I=2$\label{fig:firstm} } }
	\qquad
	\parbox{0.48\textwidth}{ \centering
		\includegraphics[width=.48\textwidth]
		  {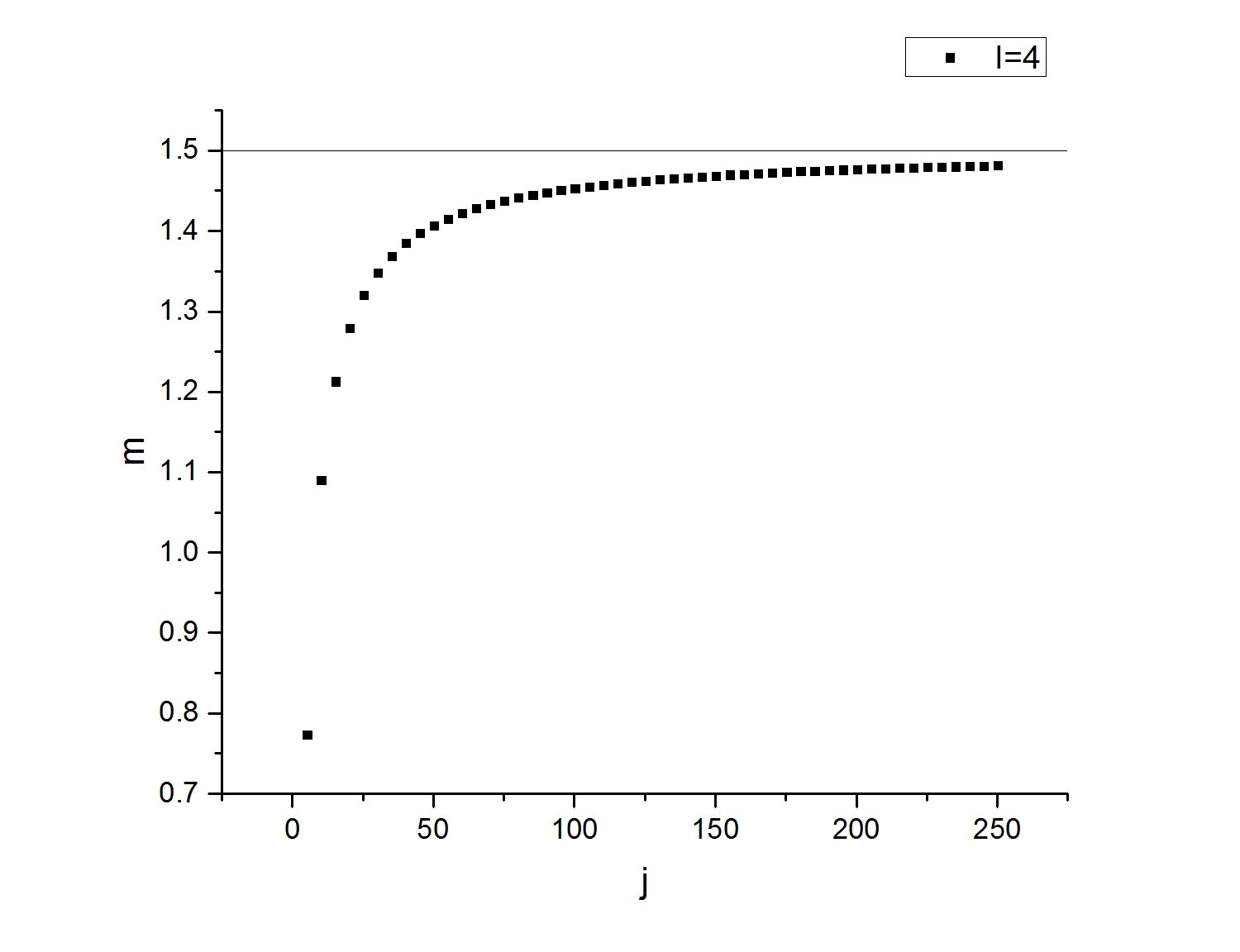}
		\caption{Suspected $m$ vs $j$; $I=4$} } 
\end{figure}

\begin{figure}
     \centering
	\parbox{0.48\textwidth}{ \centering
		\includegraphics[width=.48\textwidth]
		  {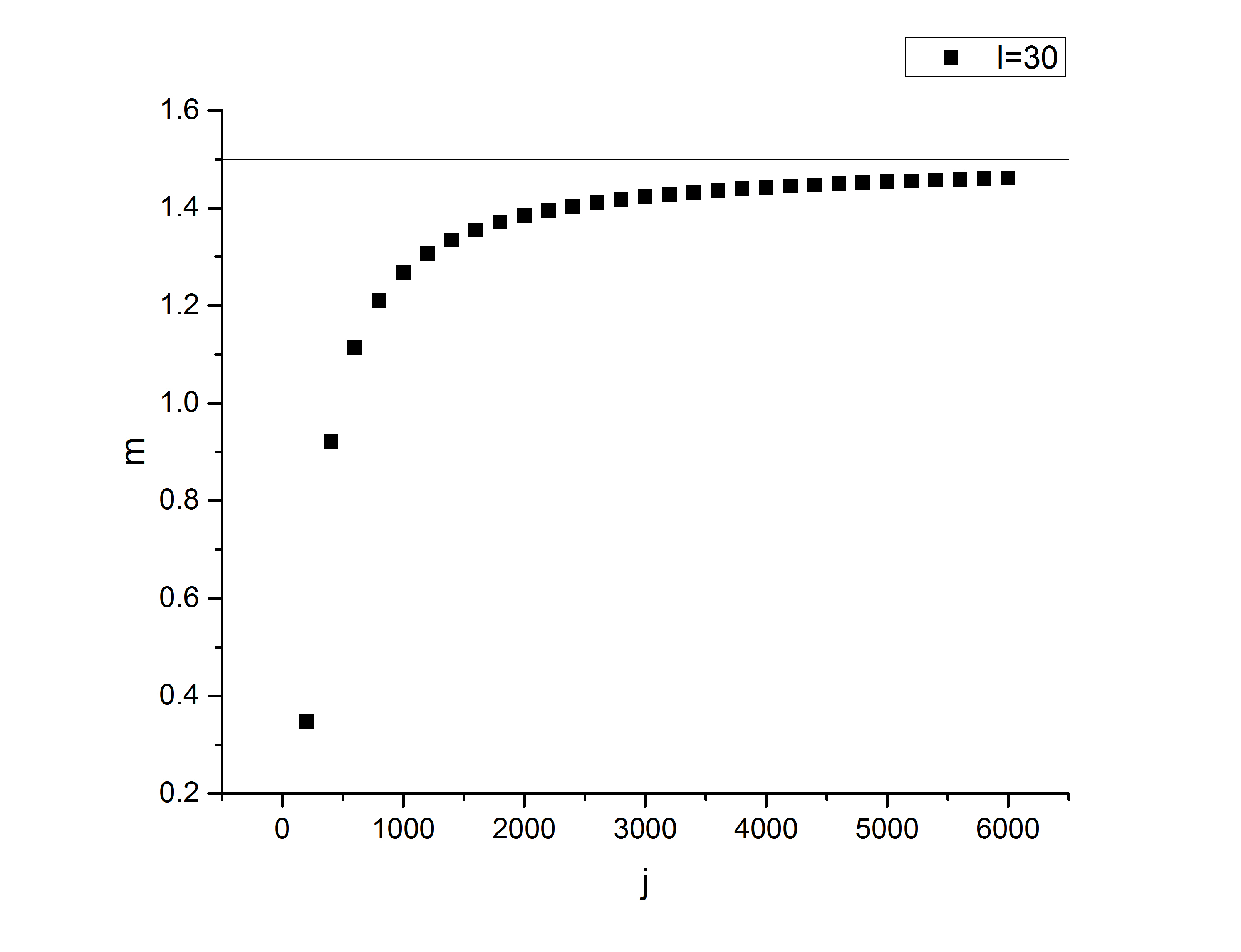}
		\caption{Suspected $m$ vs $j$; $I=30$} }
	\qquad
	\parbox{0.48\textwidth}{ \centering
		\includegraphics[width=.48\textwidth]
		  {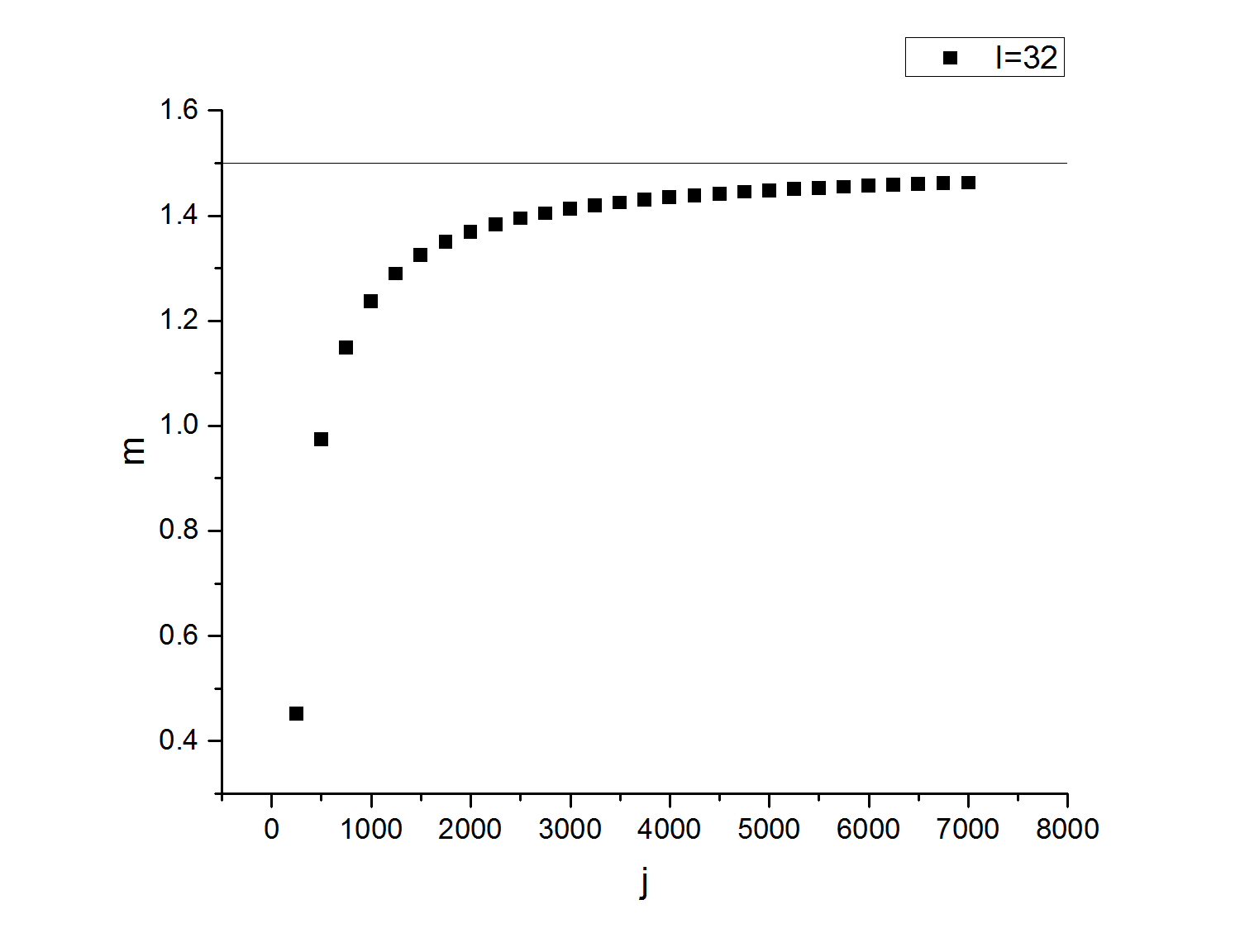}
		\caption{Suspected $m$ vs $j$; $I=32$
		\label{fig:lastm} } } 
\end{figure}

	We try a more elaborate form 
	\begin{equation}
	 UA(I,j) = 
	A j^m e^{\text{-}\alpha j} 
		\label{eq:asymp}
	\end{equation}
	We consider the ratio 
	\begin{equation}
		RR=\dfrac{ \text{U}(I,j+1)^2 }
		{ \text{U} (I,j) \text{\space} \text{U}(I,j+2)}
	\end{equation} 
	and compare this with 
	\begin{equation}
		RRA=\dfrac{(j+1)^{2m}}
		{(j(j+2))^m}
	\end{equation}
	We therefor have for the extracted $m$
	\begin{equation}
		m = \dfrac{ \ln(RR) }
		{\ln (RRA)}
	\end{equation}
	It should be noted that in the large $j$ limit $RRA$ approaches $1 + \dfrac{m}{j^2}$.
	We plot some selections of $m$ vs. $j$ in the attached Figures~\ref{fig:firstm} to~\ref{fig:lastm}.
	We find that all even $I$ from $I=2$ to $I=12$, $m$ converges to 1.5 in the large $j$ limit.
	
	It is important to note that in order to obtain the asymptotic value of $m$ in Eq.(\ref{eq:asymp}) one must go to a sufficiently large value of $j$.
	Furthermore the bigger the value of $I$ the higher one has to go in $j$.
	To show the perils of chasing the maximum j too small suppose we choose it to by 500.5, which a priori most would consider to be a very large number.
	The values of $m$ for $I=2,4,10,20,30$ respectively 1.495, 1.481, 1.391, 1.085, and 0.577.
	We now see a steady decrease in $m$ as $I$ increases, which could lead to the false conclusion that there is a different asymptotic value of $m$ for each $I$.
	However when we choose $j$ large enough e.g. up to 7000.5 for $I=32$ we see that the asymptotic value of $m$ is the same for all even $I$ up to $I=32$, namely $m=1.5$.
	
\begin{figure}
     \centering
	\parbox{0.48\textwidth}{ \centering
		\includegraphics[width=.48\textwidth]
		  {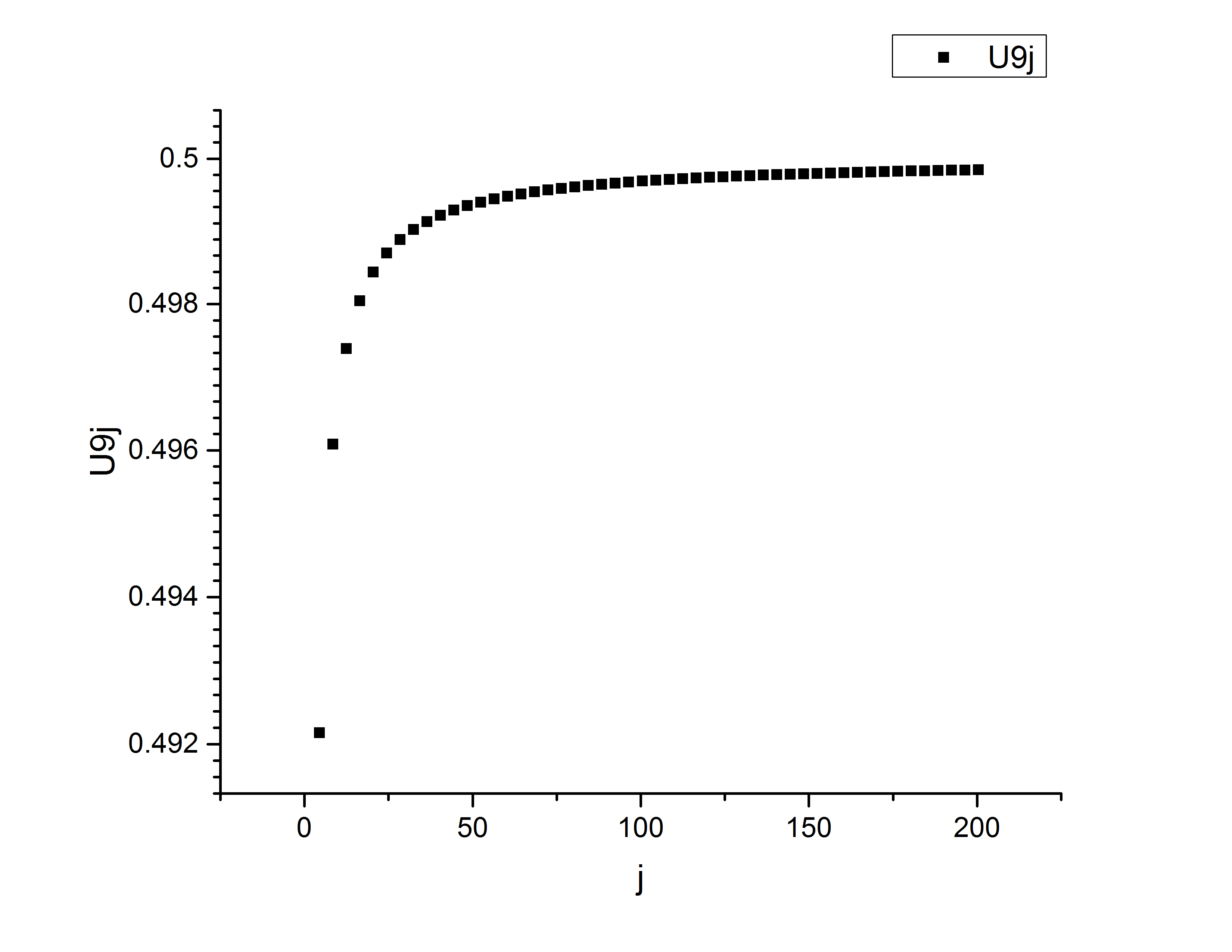}
		\caption{$U9j$(log scale) vs $j$
		\label{fig:Imax} } }
	\qquad
	\parbox{0.48\textwidth}{ \centering
		\includegraphics[width=.48\textwidth]
		  {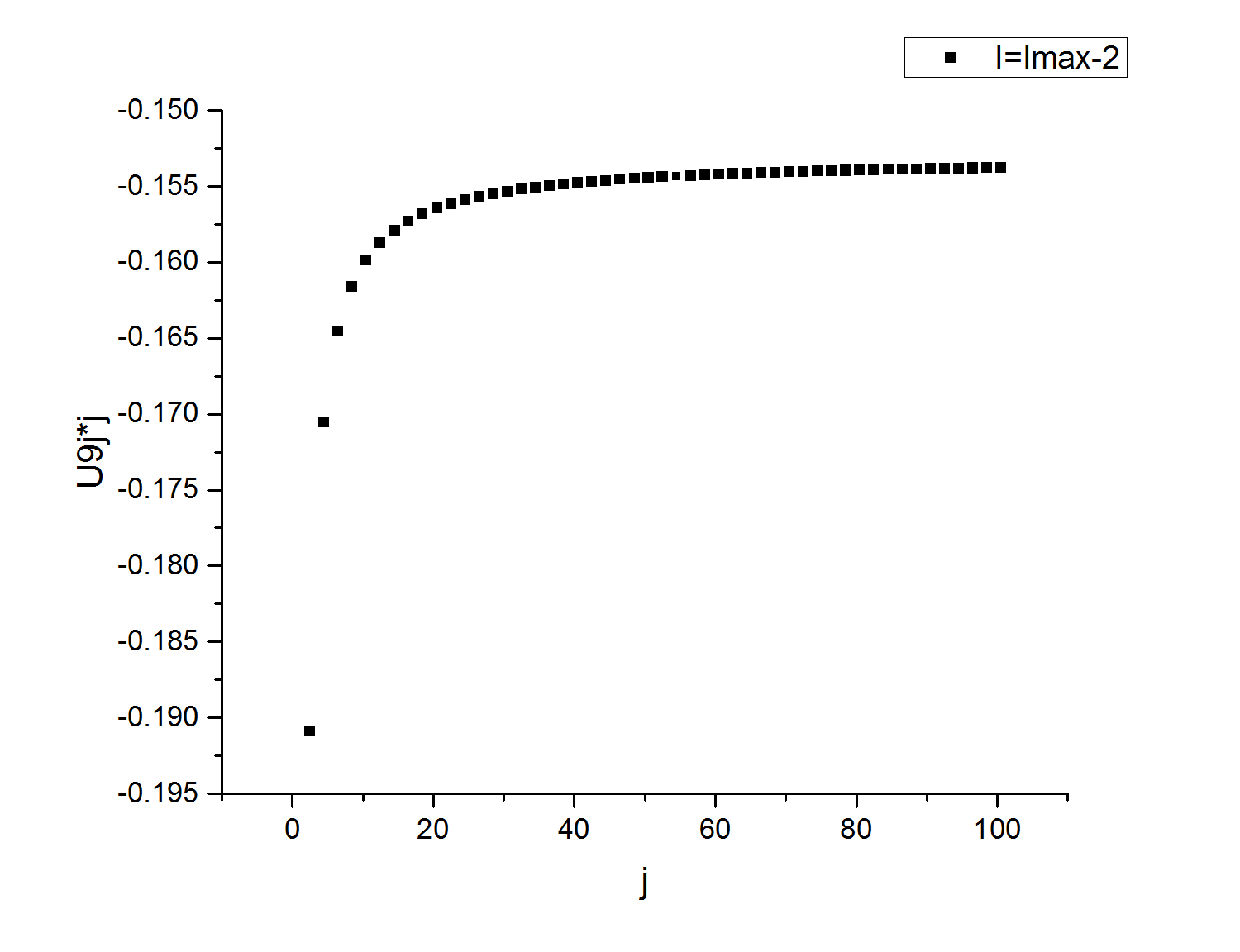}
		\caption{$U9j*j$ vs $j$} 
		\label{fig:firstImax} }
\end{figure}

\begin{figure}
     \centering
	\parbox{0.48\textwidth}{ \centering
		\includegraphics[width=.48\textwidth]
		  {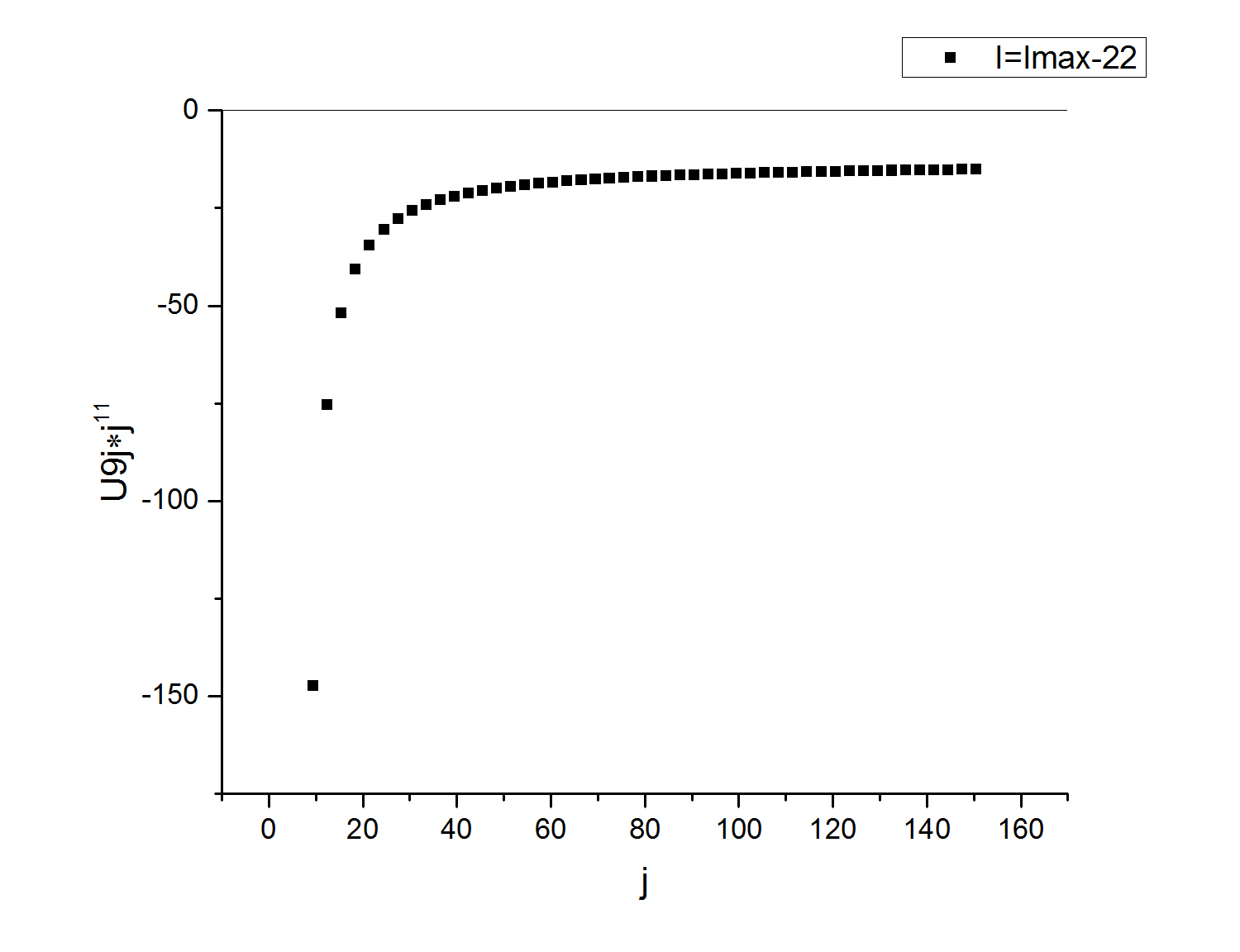}
		\caption{$U9j*j^{11}$ vs $j$} }
	\qquad
	\parbox{0.48\textwidth}{ \centering
		\includegraphics[width=.48\textwidth]
		  {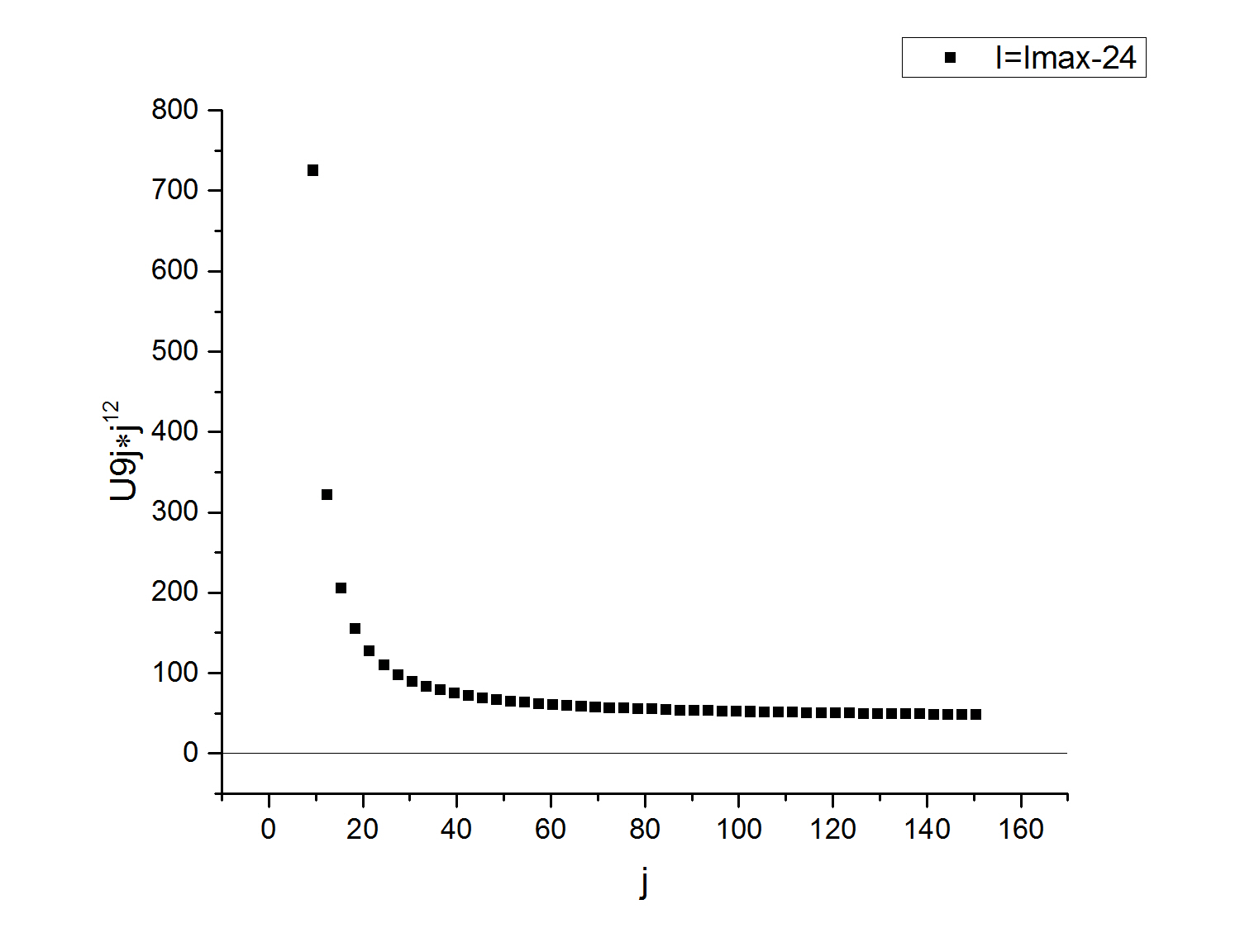}
		\caption{$U9j*j^{12}$ vs $j$
		\label{fig:lastImax} } }  
\end{figure}

	We next consider $U(I,j)$ for the largest values of $I$.
	We start with $I=I_{\text{max}}=4j - 2$ and then also consider $I_{\text{max}}-2$, $I_{\text{max}}-4$, etc.
	We find that $U(I_{\text{max}},j)$ approached a constant for large $j$.
	Then up to a certain point $U(I,j)$ drops off as $\dfrac{1}{j^n}$ where $n=\dfrac{(I_{\text{max}}-I)}{2}$ e.g. as $\dfrac{1}{j}$ for $I=4j-4$ and $\dfrac{1}{j^2}$ for $I=4j-6$.
	Asymptotic decreasing exponential behaviors for $6j$ coefficients have been presented by Varshalovich et al. in section 9.9 \cite{3}.
	The results are shown in Figures~\ref{fig:firstImax} to~\ref{fig:lastImax}.
	
	Igal Talmi in \cite{4} has an analytic formula for the case $I=I_{\text{max}}=4j-2$ (Figure~\ref{fig:Imax}) as shown here:
	\begin{equation}
		{\Big<(jj)^{2j} (jj)^{2j} \Big| (jj)^{2j} (jj)^{(2j - 2)} \Big>}^I=
		\dfrac{\sqrt{(2j-1)(8j-1)}} {2(4j-1)}
	\end{equation}
	with $I=4j-2$. Talmi has also shown that for the case $I=I_{\text{max}}-2$ (Figure~\ref{fig:firstImax}), $U9j$ goes asymptotically to $\dfrac{-1}{8j}$.
	
	A formula involving many factorials for the case $I=I_{\text{max}}$ is also given by Varshelovich et al. in sec.10:8:4 Eq.(14) in \cite{4}.
	We finally remind the reader that our motivation for this work comes from our desire to ether understand the wave function arising from a "maximum $J$-pairing" Hamiltonian \cite{1,2}.
	
% % % % % % % % % % % % % % % % % % % % % REFERENCES

	We would like to thank Igal Talmi for his valuable help and interest.
	\mbox{}\\\\

\newgeometry{margin=.5cm}
\pagestyle{empty}

\begin{figure}
\centering
   \begin{subfigure}{0.35\textheight} \centering
     \includegraphics[width=\textwidth]{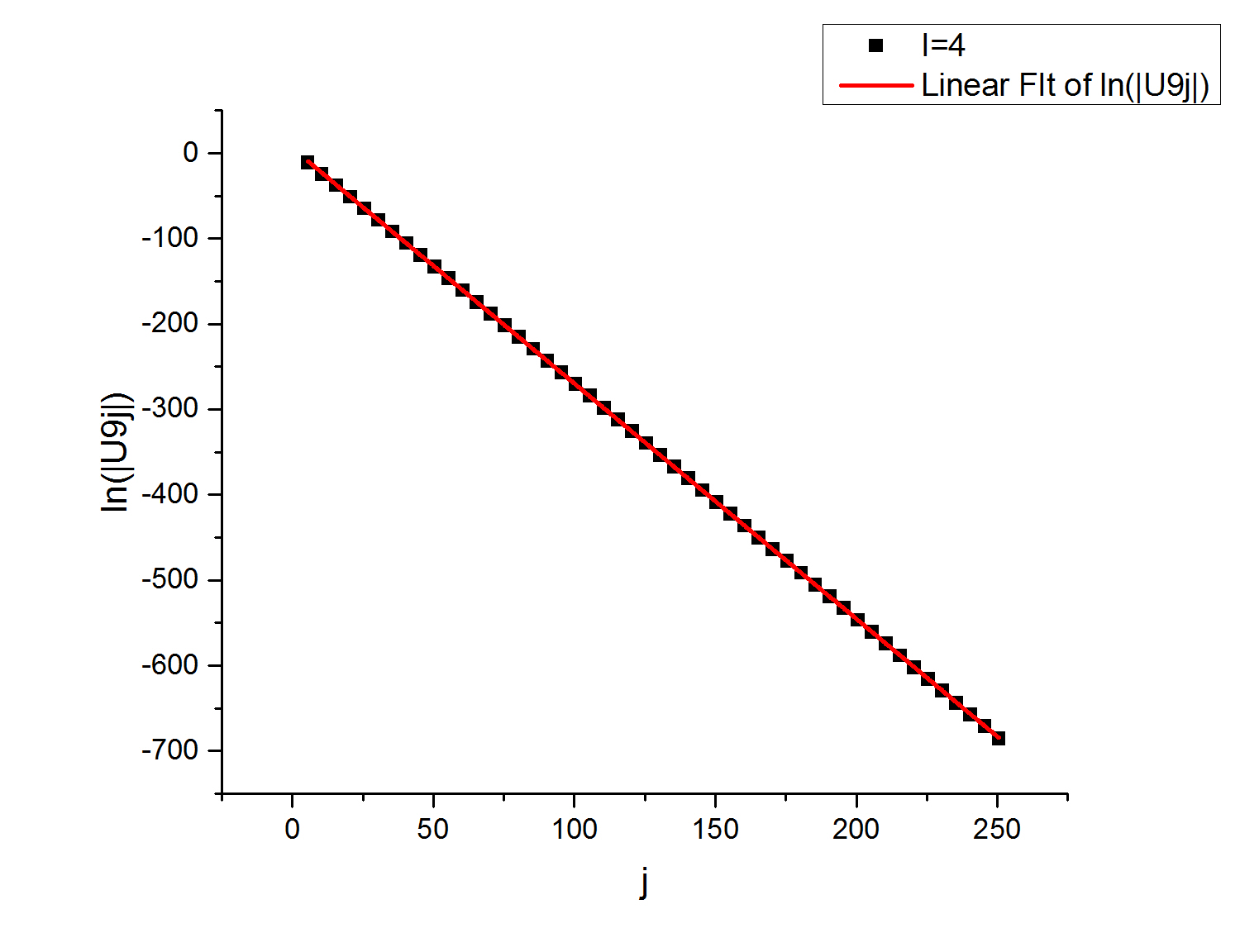}
     \caption{$\ln(|U9j|)$ vs $j$}
   \end{subfigure}
   \begin{subfigure}{0.35\textheight} \centering
     \includegraphics[width=\textwidth]{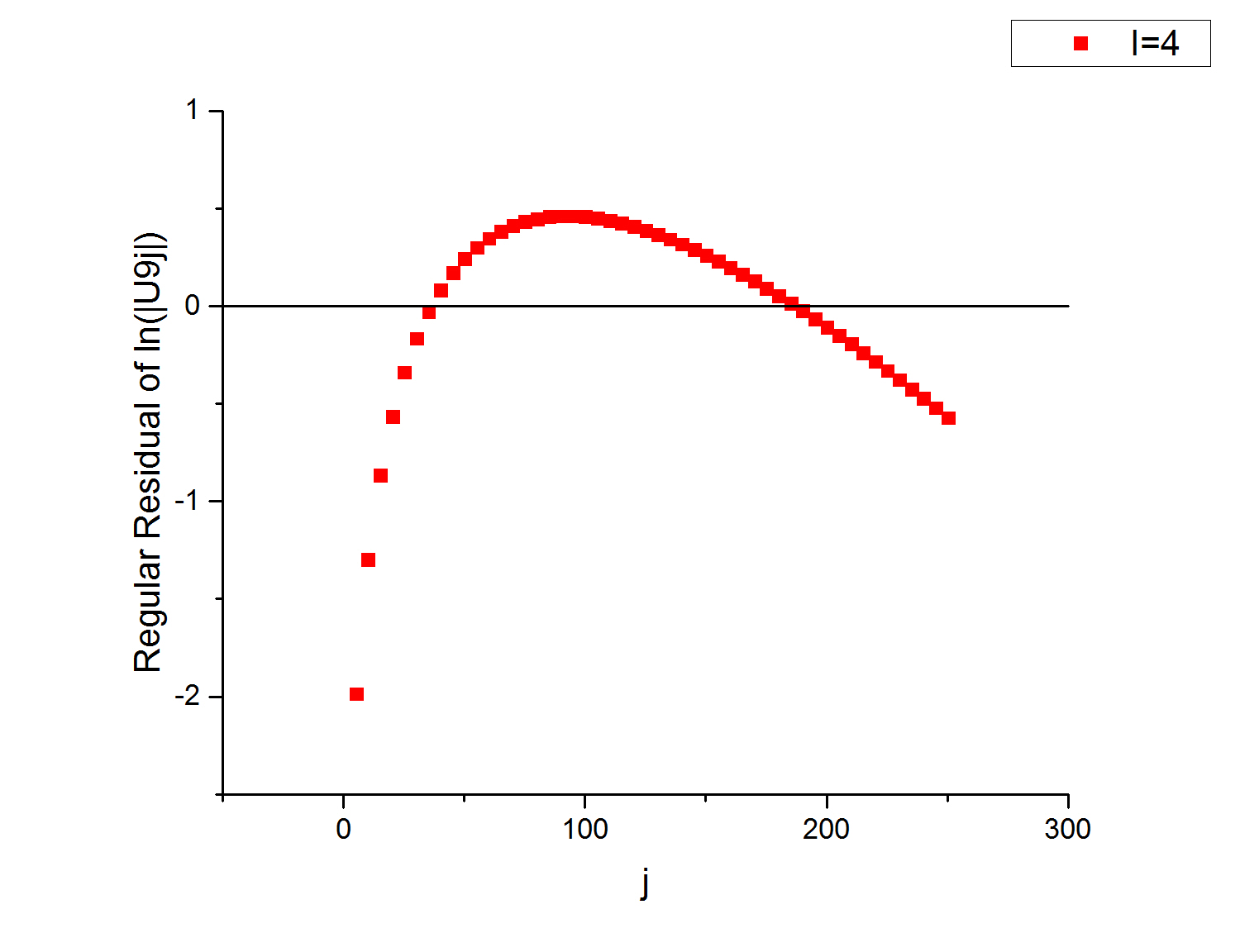}
     \caption{Residual of $\ln(|U9j|)$ vs $j$}
   \end{subfigure}
\caption{$I=4$} 
\end{figure}
\begin{figure}
\centering
   \begin{subfigure}{0.35\textheight} \centering
     \includegraphics[width=\textwidth]{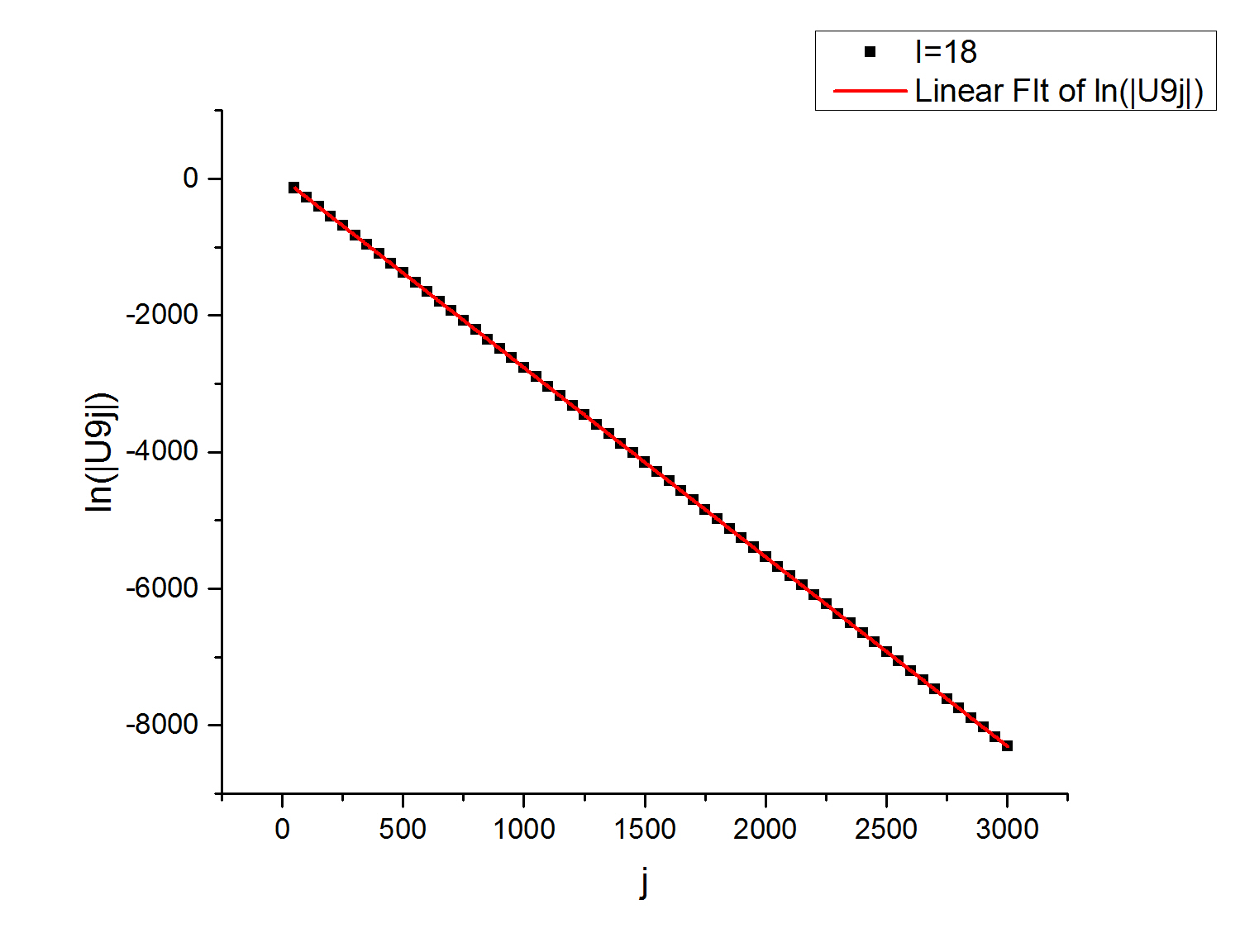}
     \caption{$\ln(|U9j|)$ vs $j$}
   \end{subfigure}
   \begin{subfigure}{0.35\textheight} \centering
     \includegraphics[width=\textwidth]{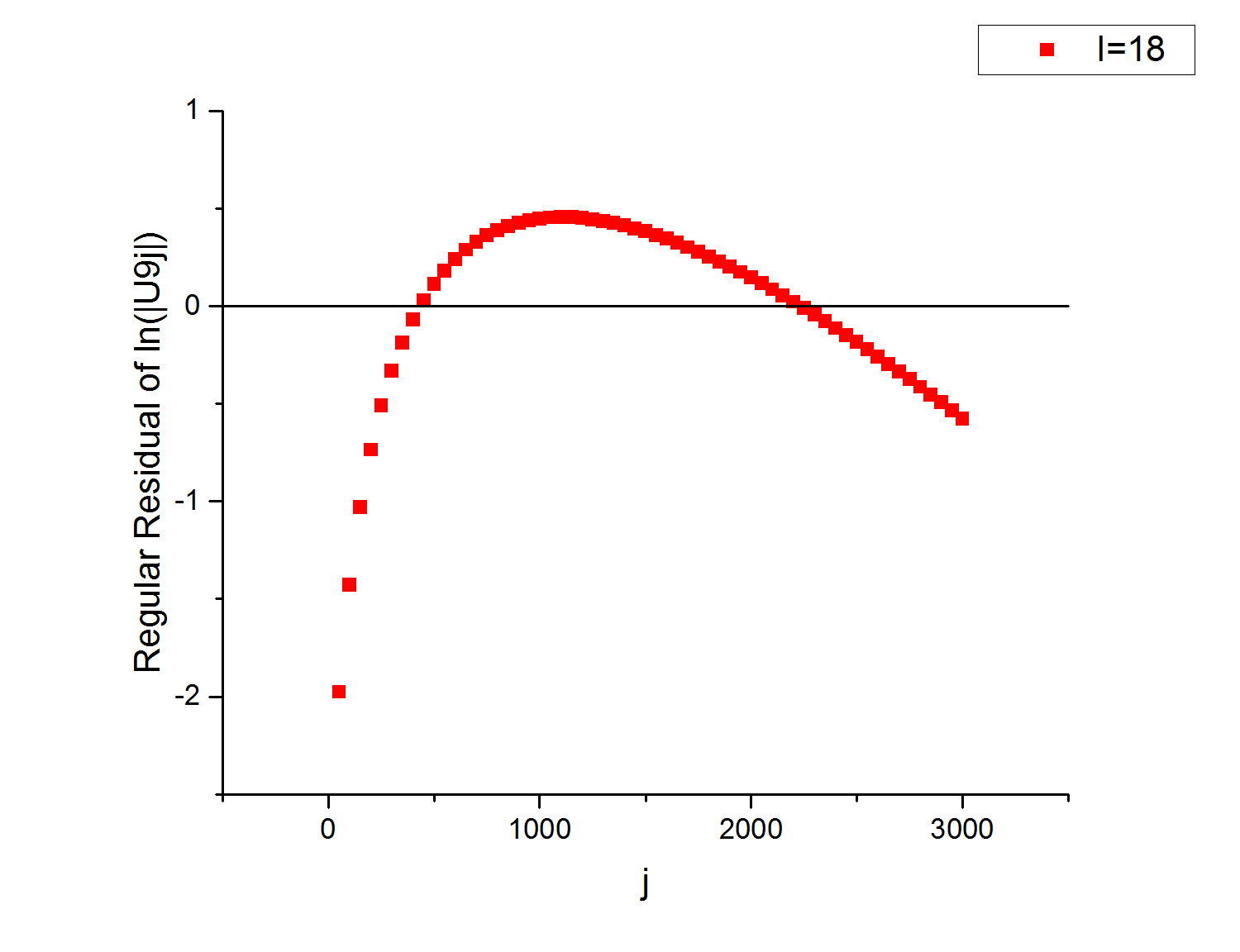}
     \caption{Residual of $\ln(|U9j|)$ vs $j$}
   \end{subfigure}
\caption{$I=18$} 
\end{figure}

\begin{figure}
\centering
   \begin{subfigure}{0.32\textheight} \centering
     \includegraphics[width=\textwidth]{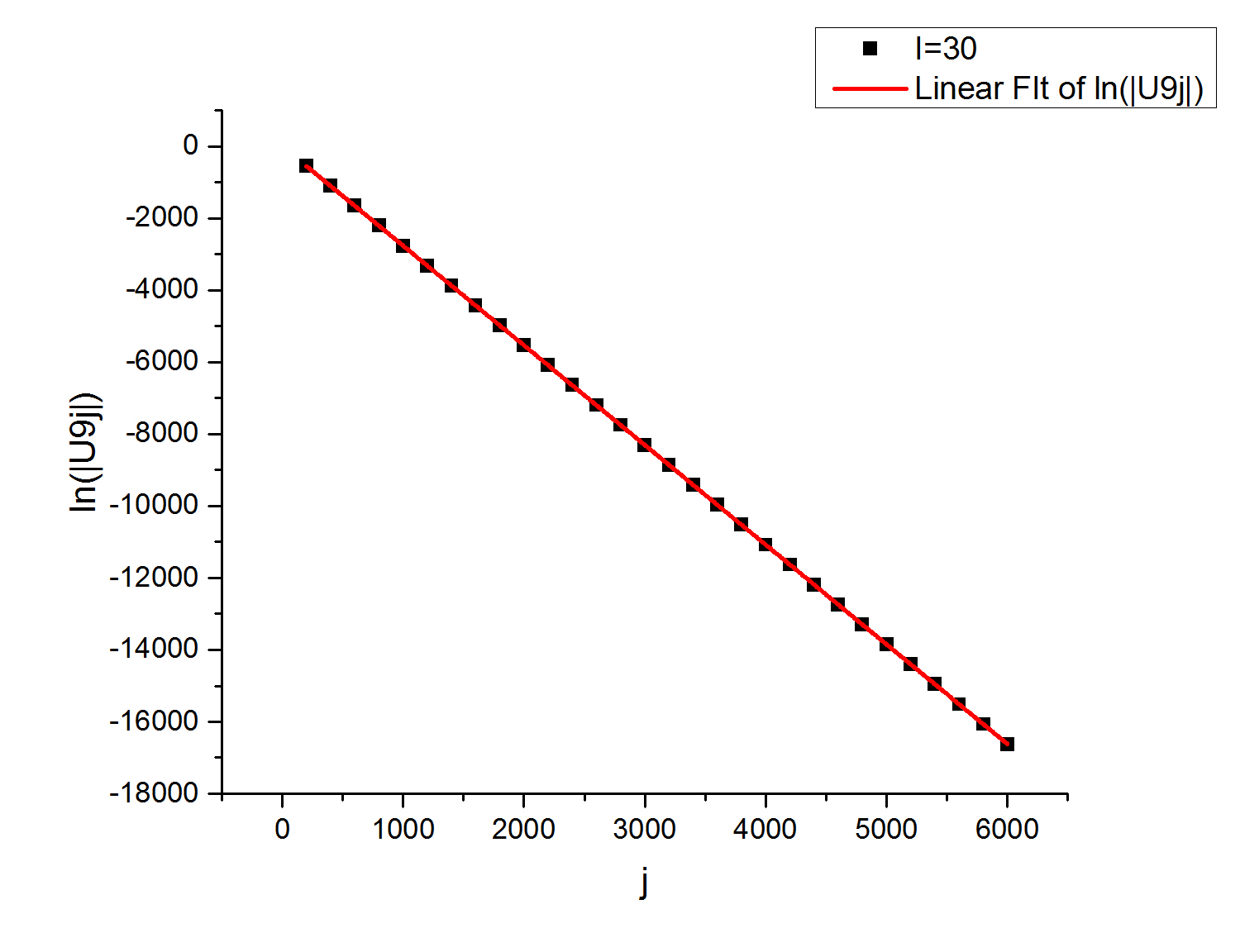}
     \caption{$\ln(|U9j|)$ vs $j$}
   \end{subfigure}
   \begin{subfigure}{0.35\textheight} \centering
     \includegraphics[width=\textwidth]{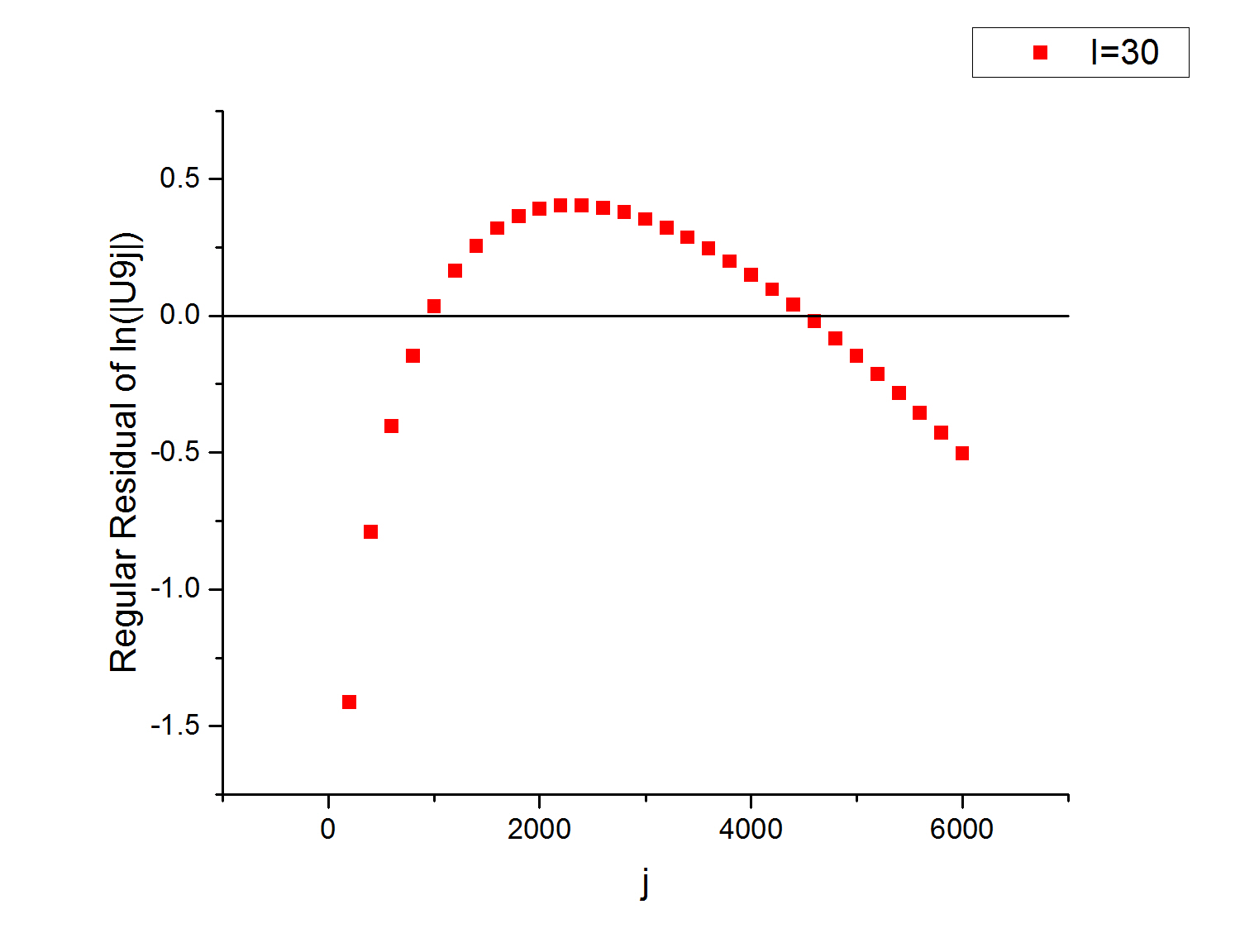}
     \caption{Residual of $\ln(|U9j|)$ vs $j$}
   \end{subfigure}
\caption{$I=30$} 
\end{figure}

% % % % % % % % % % % % % % % % % % % %
% % % % % % % % % % % % % % % % % % % % 
% % % % % % % % % % % % % % % % % % % %

\begin{figure}
     \centering
	\parbox{0.48\textwidth}{ \centering
		\includegraphics[width=.48\textwidth]
		  {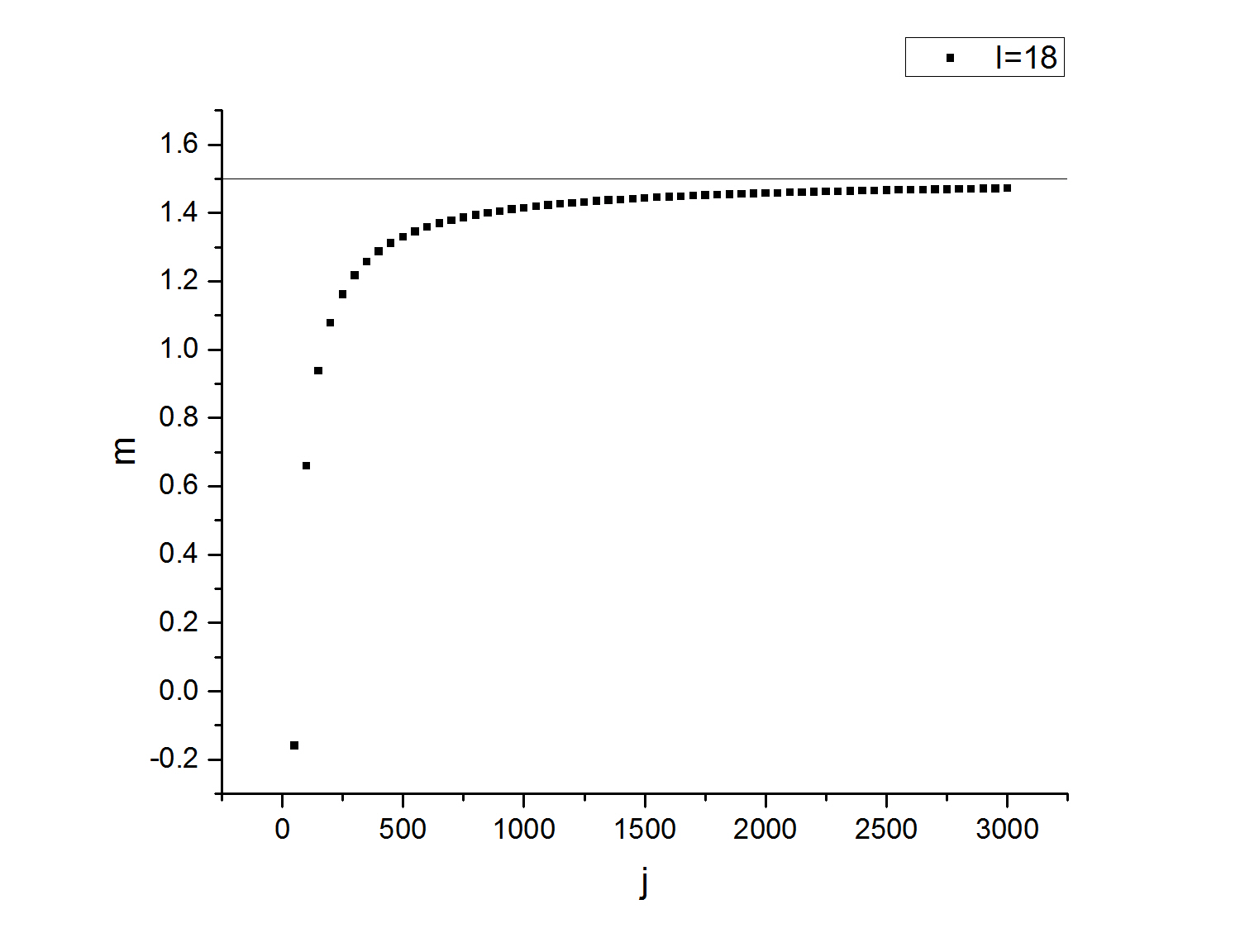}
		\caption{Suspected $m$ vs $j$; $I=18$} }
	\qquad
	\parbox{0.48\textwidth}{ \centering
		\includegraphics[width=.48\textwidth]
		  {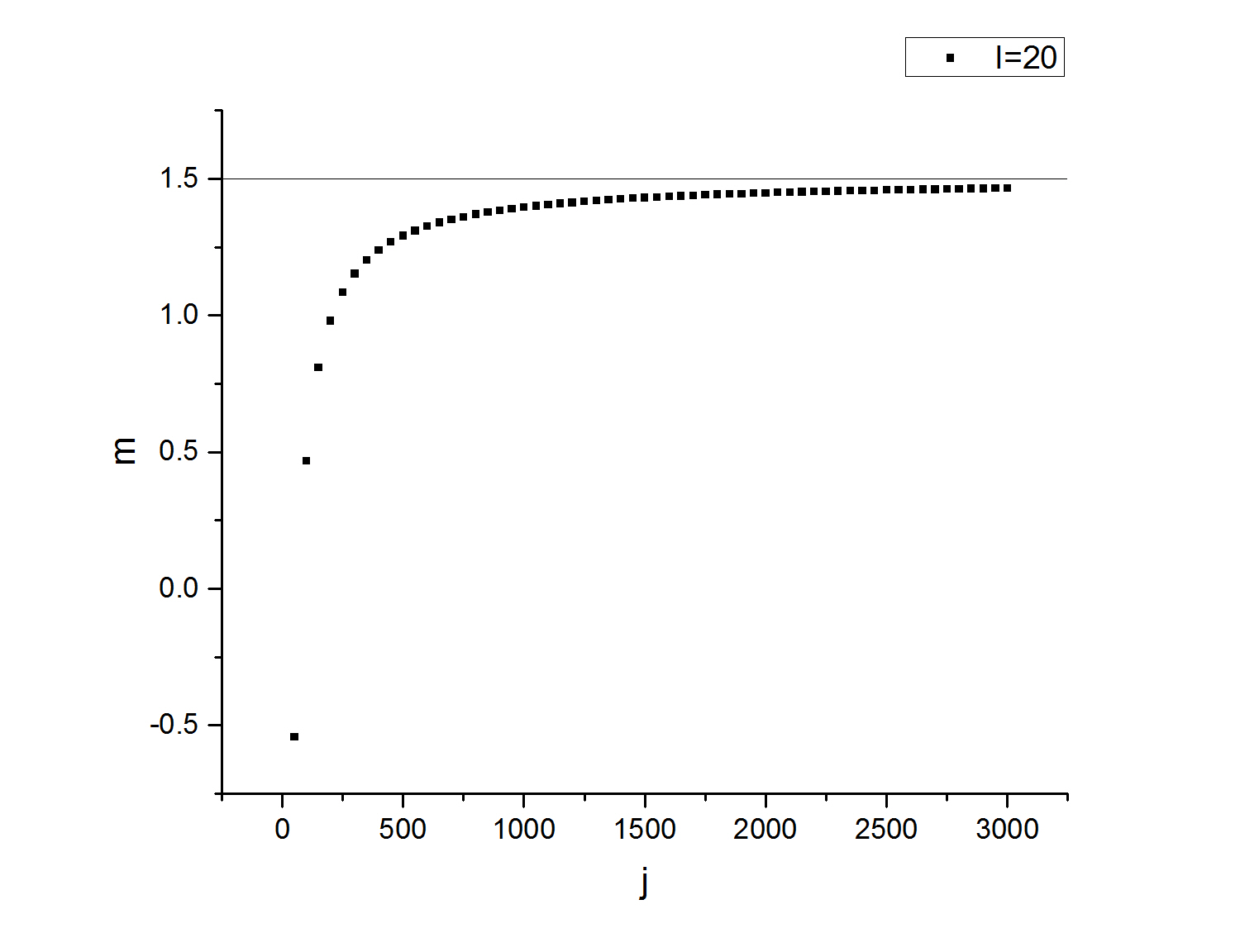}
		\caption{Suspected $m$ vs $j$; $I=20$} } 
\end{figure}

\begin{figure}
     \centering
	\parbox{0.48\textwidth}{ \centering
		\includegraphics[width=.48\textwidth]
		  {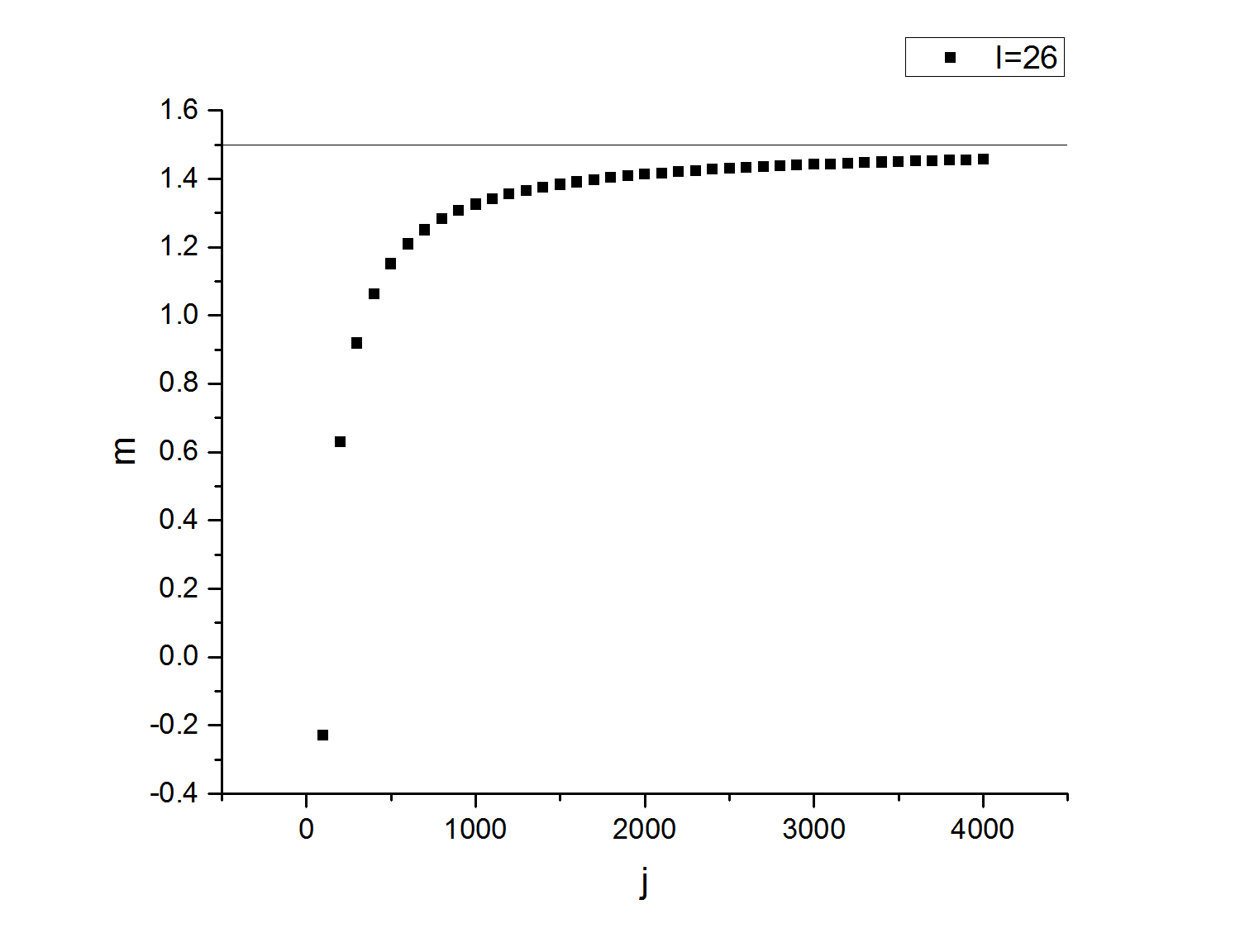}
		\caption{Suspected $m$ vs $j$; $I=26$} }
	\qquad
	\parbox{0.48\textwidth}{ \centering
		\includegraphics[width=.48\textwidth]
		  {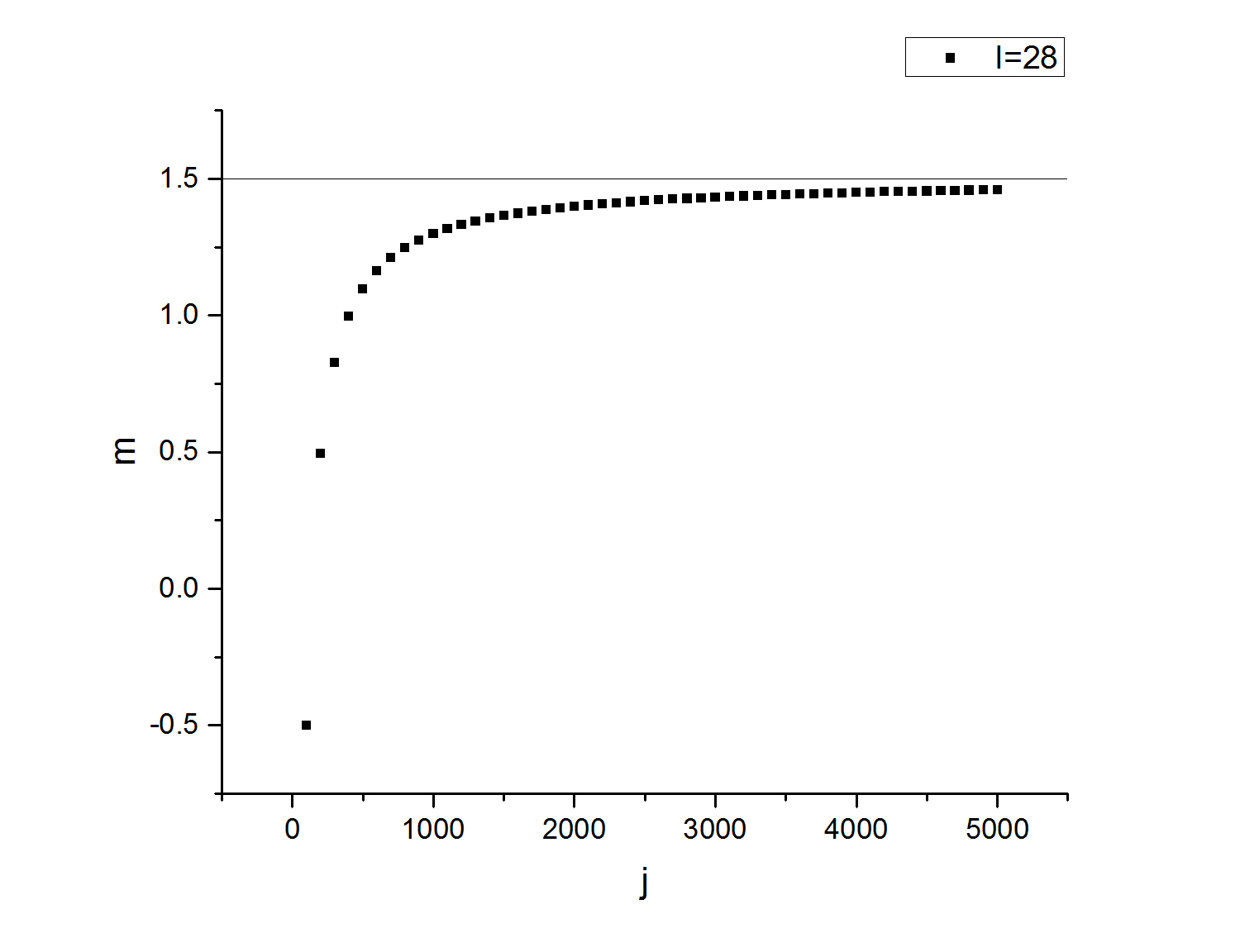}
		\caption{Suspected $m$ vs $j$; $I=28$} } 
\end{figure}

% % % % % % % % % % % % % % % % % % % % % % % % % % 
% % % % % % % % % % % % % % % % % % % % % % % % % % I=I\text{max} - 2n graphs
% % % % % % % % % % % % % % % % % % % % % % % % % % 

\begin{figure}
     \centering
	\parbox{0.48\textwidth}{ \centering
		\includegraphics[width=.48\textwidth]
		  {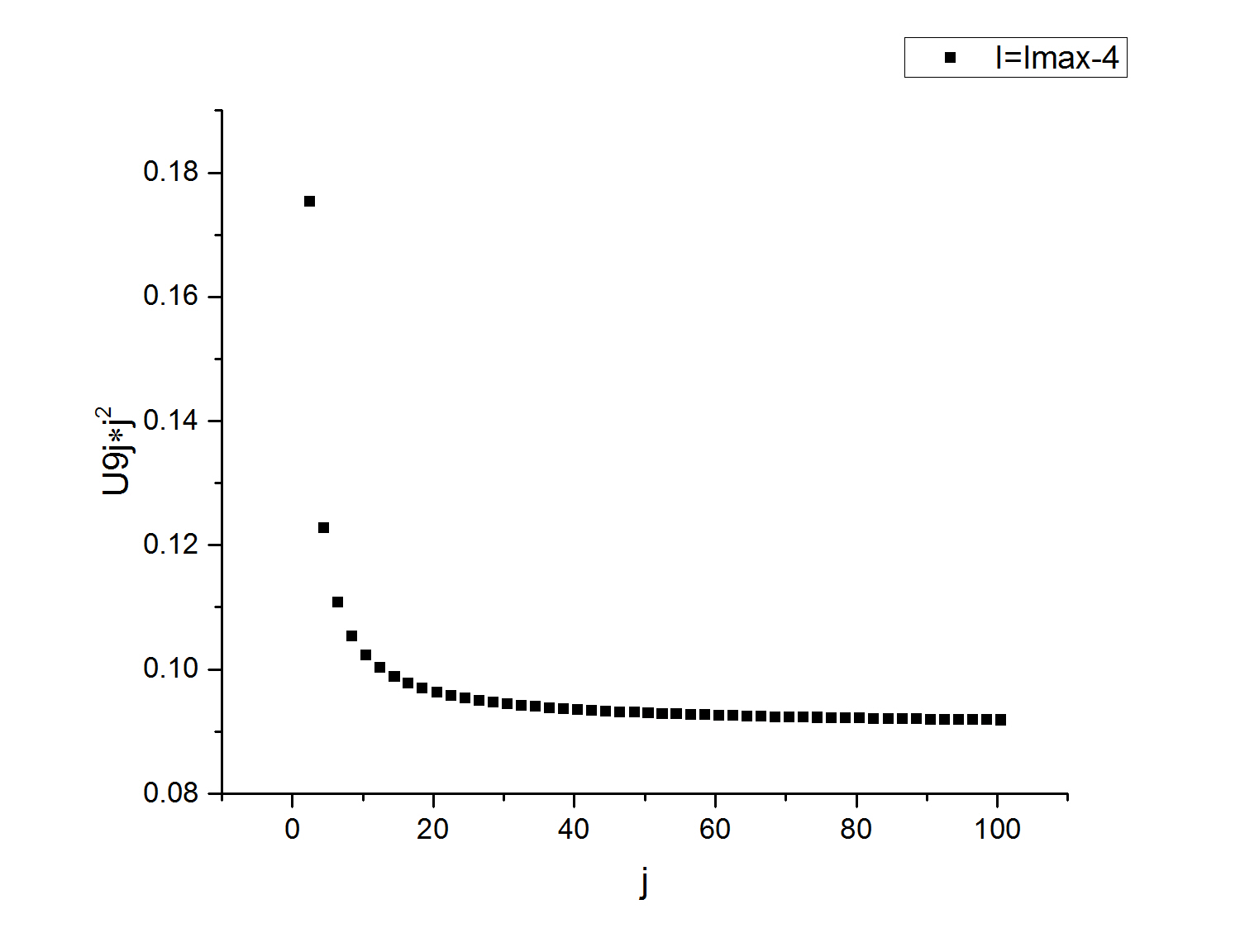}
		\caption{$U9j*j^2$ vs $j$} }
	\qquad
	\parbox{0.48\textwidth}{ \centering
		\includegraphics[width=.48\textwidth]
		  {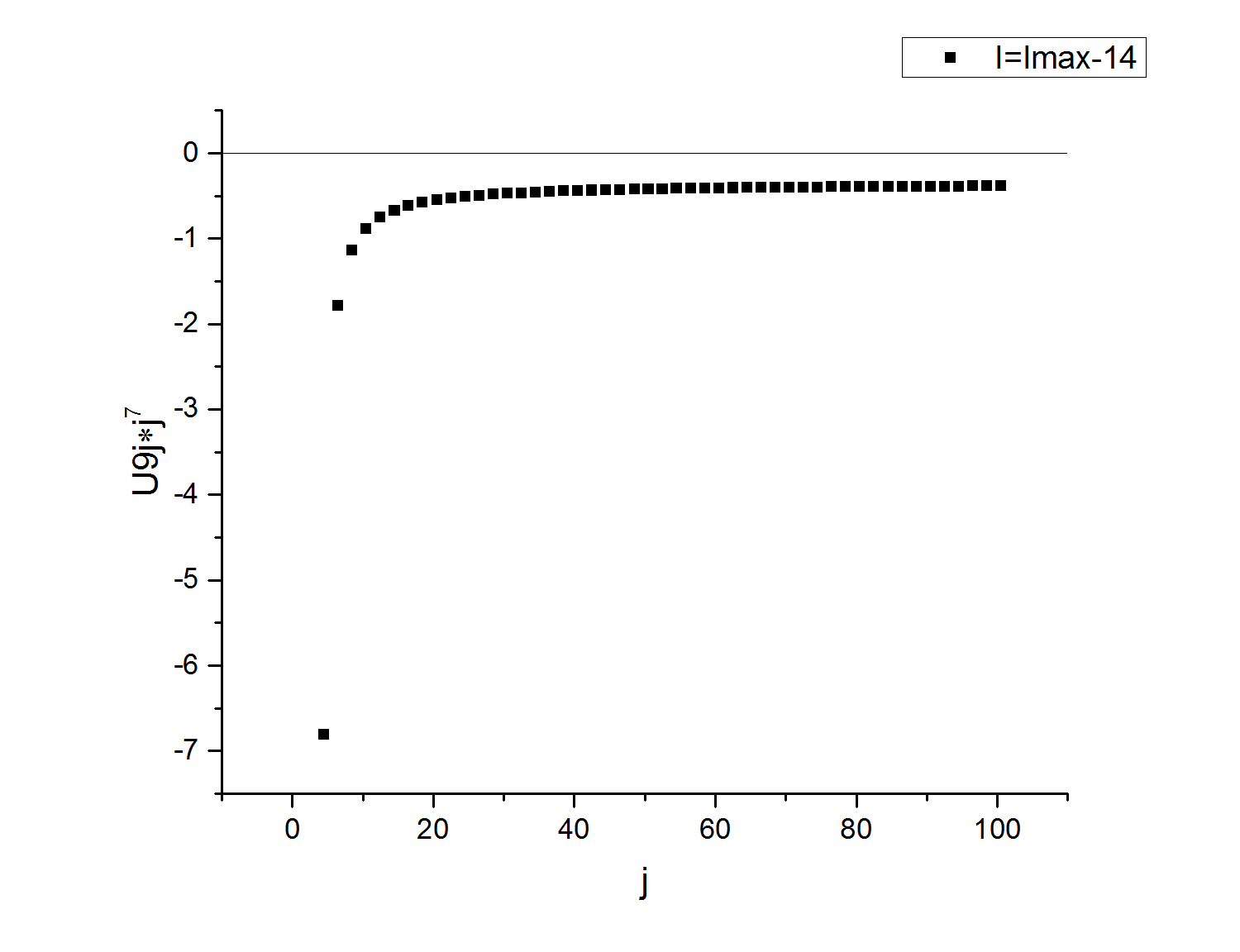}
		\caption{$U9j*j^7$ vs $j$} } 
\end{figure}

\end{document}